\documentclass[12pt,preprint]{aastex}

\begin{document}


\title{Is the CMB asymmetry due to the kinematic dipole?}


\author{P. Naselsky, W. Zhao, J. Kim and S. Chen}
\affil{Niels Bohr Institute and DISCOVERY Center, Blegdamsvej 17, 2100 Copenhagen, {\O},  Denmark }







\begin{abstract}
Parity violation found in the Cosmic Microwave Background (CMB) radiation is a crucial clue for the non-standard cosmological model 
or the possible contamination of various foreground residuals and/or calibration of the CMB data sets. In this paper, we study 
the directional properties of the CMB parity asymmetry by excluding the $m=0$ modes in the definition of parity parameters. 
We find that the preferred directions of the parity parameters coincide with the CMB kinematic dipole, 
which implies that the CMB parity asymmetry may be connected with the possible contamination of the residual dipole component. 
We also find that such tendency is not only localized at $l=2,3$, but in the extended multipole ranges up to $l\sim 22$.
\end{abstract}






\section{Introduction}

Symmetry of the physical process in our Universe and particular mechanisms of its violation is golden mind of the modern physics. 
Since pioneering Lee and Yang investigations of the parity symmetry in the weak interaction, the principle of symmetry is deeply incorporated into the modern particle physics, 
including the Higgs mechanism of symmetry breaking, in chemistry, in physics of condensed matter and, in general, in the theory of the phase transition. 
Passing from the microscopic physics to the properties of the space and time at large, we have to admit that the Cosmic Microwave Background (CMB) radiation anisotropy provides invaluable 
test for the investigation of parity at the megascopic scales above the
scale of inhomogeneity $\sim100$Mpc. The problem of the parity asymmetry of the CMB has been investigated in \citep{evil,jk1,jk2,grup,hansen,maris,david}, showing significant
dominance of the power spectrum stored in the odd multipoles over the even ones. Recently, in \citep{jk3} it was shown that the odd multipole preferences tidily connected with the anomalies of the two-point correlation function, in particular, the lack of correlations at
$60^{\rm o}\le \Theta\le 180^{\rm o}$ \citep{schwarz1,copi1}, and newly discovered anomaly of the correlations at $1^{\rm o}\le \Theta\le 30^{\rm o}$ \citep{jk3}. In combination with the
widely discussed anomalies of the CMB map and the power spectrum \citep{eriksen1} (see for review \citep{wmap_anomly}), investigation of the origin of these anomalies could put a new
light on the physics of the early Universe, the methods of the foregrounds reductions and calibration of the CMB data sets, improving our knowledge of the most fundamental cosmological parameters and  the theory of inflation.

The local motion of an observer through the CMB frame produces the so-called Kinematic Dipole (KD) anisotropy of the CMB, which is the most powerful component of the signal and fitted out from the CMB data before cosmological analysis. 
In this paper, we are going to show that some of the discovered features of the Wilkinson Microwave Anisotropy Probe (WMAP) CMB TT anisotropy, 
including the odd-parity preference of the power spectrum, could have common origin associated with the KD of the CMB. Previously, the possible contamination of the CMB by
KD has been assessed by multipole vector statistic (alignment of the quadrupole and octupole components) \citep{gordon,peiris}. 
Additionally, we will show that the low multipole anomalies are associated with other anomalies such as the lack of angular correlation, 
the even/odd-parity asymmetry and the planarity of the multipole $l=5$.
First, we will focus on the properties of the CMB TT correlation function $C(\Theta)=\Delta T(\hat{\mathbf n})\Delta T(\hat{\mathbf n'})$ at angle 
$\Theta=\arccos(\hat{\mathbf n}\cdot\hat{\mathbf n'} )=\pi$ and show
that $C(\Theta=\pi)$ is connected to the parity parameter $g(l)$. 
We will estimate the power spectrum without the $m=0$ mode so that the rotational invariance of the angular power spectrum may not be automatically satisfied.
As well known, these $m=0$ modes pick up a certain direction (i.e. the $z$-axis direction in the spherical coordinate system where $a_{lm}$ are defined \citep{gordon}). 
Using these estimators, we will investigate the possible preferred direction in the CMB field. 
If our Universe is, indeed, statistically homogeneous and isotropic with Gaussian seed perturbations, we should have no or little parity preference, given for our estimators associated with 
the angular power spectrum. 
Using the WMAP 7-yr Internal Linear Combination (ILC7) map, we compute the parity parameters for coordinates of various orientations. 
We find some level of alignment between the KD direction and the orientation, in which the parity asymmetry is greatest. 
Therefore, the CMB parity asymmetry may be related to the systematics associated with KD, which may be also responsible for the alignment problem of quadrupole and octupole \citep{gordon,dt2}.

The outline of the paper is the following. In Section 2 we introduce the basic characteristics of the CMB parity asymmetry.
In Section 3, we investigate the orientations of maximum parity asymmetry and compare them with the CMB kinematic dipole. 
In Section 4, we summarize our investigation.

\section{Odd-multipole preference of the CMB power spectrum}
The temperature fluctuations of CMB anisotropy, can be conveniently decomposed as follows:
\begin{eqnarray}\label{a_lm}
\Delta T (\theta,\phi) = \sum ^{\infty}_{l=0} \sum ^{l}_{m=-l} a_{lm} Y_{lm}(\theta,\phi),
\end{eqnarray}
where $a_{lm}$ are the coefficients of decomposition: $a_{lm} = |a_{lm}| \exp(i \phi_{lm})$, with $\phi_{lm}$ as the phase.
Under the assumption of total Gaussian randomness, as predicted by the large class of inflationary models, the amplitudes $|a_{lm}|$ are distributed according to Rayleigh's probability distribution function and the phases of $a_{lm}$ are supposed to be evenly distributed in the range $[0,2 \pi]$ \citep{bardeen1986}.

\begin{figure}
\epsscale{1.00}
\plotone{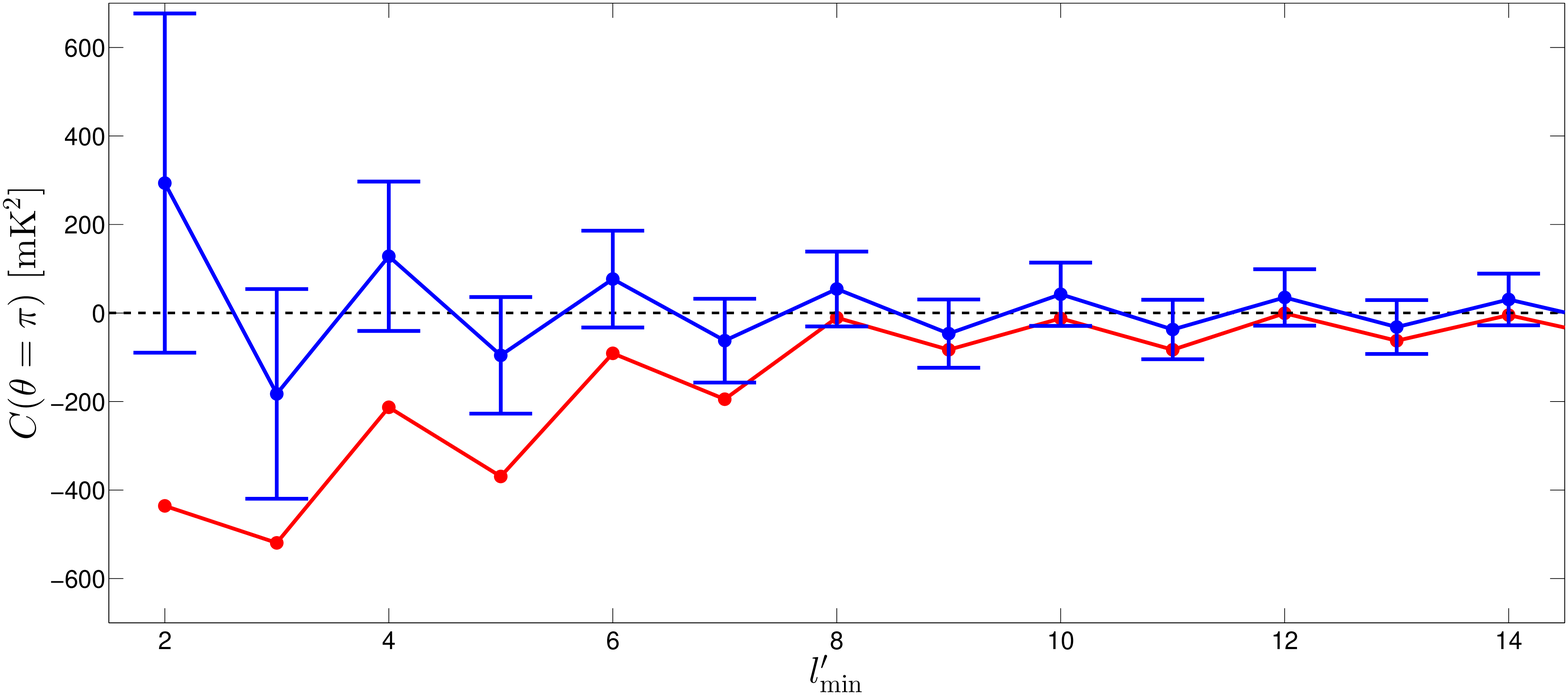}
\caption{The theoretical (blue curve) and observed (red curve) values of $C(\theta=\pi)$ as a function of $l'_{\rm min}$, 
where the WMAP7 power spectra $C(l)$ have been used as the observed data. 
The error bars indicate the $1\sigma$ confident levels caused by cosmic variance. }.\label{com2}
\end{figure}

For any signals $T(\hat{\mathbf n})$ defined on the sphere, one can extract symmetric ($ \Delta T^+(\hat{\mathbf n})=\Delta T^+(-\hat{\mathbf n} $)) and antisymmetric ($ \Delta T^-(\hat{\mathbf n}) =- \Delta T^-(-\hat{\mathbf n})$) components,
where
\begin{eqnarray}
\Delta T^{\pm}(\hat{\mathbf n})&=&\sum_l\sum_{m=-l}^{l}a_{lm}\Gamma^{\pm}(l)Y_{lm}(\hat{\mathbf n}),
\end{eqnarray}
and $\Gamma^{+}(l)\equiv\cos^2(\frac{\pi l}{2})$, $\Gamma^{-}(l)\equiv\sin^2(\frac{\pi l}{2})$, $Y_{lm}(\hat{\mathbf n})=(-1)^l\,Y_{lm}(-\hat{\mathbf n})$. Naive expectation, where
 the concordant $\Lambda$CDM cosmological model with initial statistically isotropic and Gaussian adiabatic perturbations is assumed, is the absence of any features distinct between even and odd multipoles. However, in reality this statement needs more accurate clarification. In particular, for the curvature perturbations beyond the present horizon the power spectrum
is given by $P(k)\propto k^{-4+n_s}$, where $n_s\simeq 0.96$ is the spectral index of the primordial density perturbations \citep{komatsu2011}. Thus, the
variance of the metric perturbations $\sigma^2\sim\int k^2P(k)dk\propto k_{\min}^{n_s-1}$ has very weak power-law ($n_s<1$) or logarithmic divergence ($n_s\simeq 1$), if $k_{\min}\rightarrow 0$. Since the low multipole range of the CMB temperature anisotropy is
determined by the ordinary and integrated Sachs-Wolfe effects, these peculiarity of the power spectrum of metric perturbations are crucial for the
two-point correlation function:
\begin{eqnarray}
{ C_{th}(\Theta)}&\equiv& <\Delta T(\hat{\mathbf n})\Delta T(\hat{\mathbf n'})> \nonumber \\
&=&\sum_{l=l_{\min}}^{\infty}\frac{2l+1}{4\pi}C_{th}(l)P_l(\cos\Theta),
\label{corf}
\end{eqnarray}
where $C_{th}(l)$ are the theoretical power spectrum, $P_l(\cos\Theta)$ are the Legendre polynomials, $\cos\Theta={\mathbf n}\cdot{\mathbf n'}$, and $<..>$ denotes the average over the statistical ensemble of realizations.
Using Eq. (\ref{corf}), we may easily show, for the largest angular distance:
\begin{eqnarray}
{C_{th}(\Theta=\pi)}=\sum_{l=l_{\min}}^{\infty}\frac{2l+1}{4\pi}C_{th}(l)(\Gamma^+(l)-\Gamma^-(l)).
\label{corpi}
\end{eqnarray}
As clear in Eq. (\ref{corf}), the natural way to estimate the relative contribution of even and odd multipoles to the correlation function
is to define the statistic
\begin{eqnarray}
 g(l)=\frac{\sum_{l'=l'_{\min}}^{l}\frac{2l'+1}{4\pi}C(l')\Gamma^+(l')}{\sum_{l'=l'_{\min}}^{l}\frac{2l'+1}{4\pi}C(l')\Gamma^-(l')},
\label{est}
\end{eqnarray}
where $l'_{\min}=1$ or $2$ (see discussion in the forthcoming sections).
Note that this statistic differs from the $g(l)$, widely used in \citep{jk1,jk2,jk3}, in the sense that $l'(l'+1)$ is replaced by $2l'+1$.
Then, from Eqs. (\ref{corpi}) and (\ref{est}) we get the estimator of the quantity ${C_{th}(\Theta=\pi)}$:
\begin{eqnarray}
{C(\Theta=\pi)}= P^-(l)\left[g(l)-1\right],\nonumber\\
P^{\pm}(l)= \sum_{l'=l'_{\min}}^{l}\frac{2l'+1}{4\pi}C(l')\Gamma^{\pm}(l').
\label{est1}
\end{eqnarray}
Thus, if $g(l) = 1$, the corresponding correlation function is ${C(\Theta=\pi)} = 0$. 
In reality, the theoretical correlation function shows some parity asymmetry. 
For instance, the dipole component naturally contributes the odd parity, while the quadrupole contributes the even parity.
When the background cosmological parameters correspond to the concordant $\Lambda$CDM model,
the properties of theoretical correlation function ${C_{th}(\Theta=\pi)}$ depends on the value of $l'_{\min}$, 
which is clearly shown in Fig. \ref{com2}. We find that the odd parity is perferred when the odd $l'_{\rm min}$ is chosen, 
while the even parity is perferred when the even $l'_{\rm min}$ is chosen. 
However, the observed data show the different tendency for the parity asymmetry, i.e. the parity violation comparing with the theoretical predictions. 
From Fig. \ref{com2}, we find that the odd parity is always perferred for all the cases with $l'_{\rm min}\le 15$.

\begin{figure}
\epsscale{.80}
\plotone{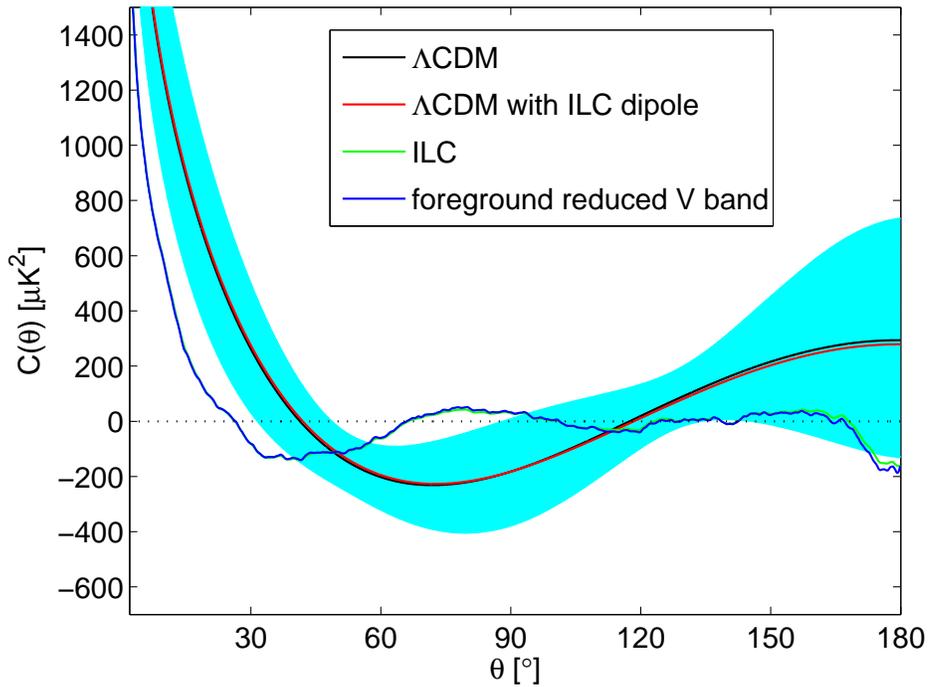}
\caption{The TT correlation function estimated from WMAP7 observational data sets and a theoretical prediction with 1$\sigma$ interval (shaded with Cyan color).
We estimated the theoretical prediction respectively with $l'_{\mathrm{min}}=2$ and $l'_{\mathrm{min}}=1$, where we used the residual dipole anisotropy of the ILC7 map.
The foreground-contaminated region in the data sets is excluded by the WMAP KQ75 mask.}.\label{corfunc}
\end{figure}

Usually, the dipole component of ${C_{th}(\Theta=\pi)}$ is not
included in Eq. (\ref{corf}), and the lower limit is set to $l'_{\min}=2$. In Fig. \ref{corfunc} we 
carefully calculate the WMAP7 observational
correlation functions by considering the KQ75 mask and a theoretical prediction with 1$\sigma$ interval, where the cosmic variance effect is taken into account. 
Consistent with Fig. \ref{com2}, we expect the theoretical value of ${C_{th}(\theta=\pi)}$ is positive, and the zero value is included at $68\%$ C.L.
Therefore the Universe has tendency to be parity asymmetric $g(l)>1$ by choosing $l'_{\rm min}=2$, with very low chance of $g(l)<1$.
However, the WMAP7 data show that ${C(\theta=\pi)}<0$ at $95\%$ C.L., and our Universe belongs to the very rare realizations
of the $\Lambda$CDM cosmological models with given by the WMAP7 cosmological parameters.

If the parity violation has the cosmological origin, we should also
see the similar parity violation in TE, EE and BB components of the polarized signal. 
However, in the curret WMAP data, the noises of the polarization data, including the TE cross-correlation data are quite large. 
So we cannot get any solid results on the parity asymmetry by using the WMAP polarization data, when taking into account the large error bars.
It is expected that the forthcoming Planck data would provide the much better chance to study the parity asymmetry in the polarization data, 
and be helpful to reveal of the origin of the parity violation.


\section{Directional statistic of the parity asymmetry}

As shown in the previous section, for the random Gaussian statistically isotropic and homogeneous
perturbations of the CMB, the correlation function $C(\Theta)$  is fully determined by the power spectrum
$C(l)$, which is rotationally invariant. Statistical invariance
means that for any rotations of the reference system of
coordinate, the power spectrum and the correlation function
are invariant. The idea of the method, proposed in this
section, is to replace the power spectrum $C(l)$ in Eq.
(\ref{corf}) by a rotationally variant power spectrum $D(l)$, defined
as
\begin{eqnarray}
 D(l)\equiv\frac{1}{2l+1}\sum_{m=-l}^l|a_{lm}|^2(1-\delta_{m0}),
\label{dpow}
\end{eqnarray}
where $\delta_{mm'}$ is the Kroneker symbol.

As it is seen from the definition given by Eq. (\ref{dpow}), the relative difference between $D(l)$ and $C(l)$ is given by $\Delta (l)\equiv \frac{D(l)-C(l)}{C(l)}=-a^2_{l0}/\sum_m|a_{lm}|^2$.
So we have $\Delta(l)\sim O(\frac{1}{2l})$ for random Gaussian CMB field. Thus, the major difference $\Delta(l)$ comes from $l=2$ and $l=3$ modes, while for $l\ge 5$ their contributions are smaller than $10\%$.

Now, we can study the power spectrum $D(l)$ in any coordinate system. Imagining the Galactic coordinate system is rotated by the Euler angle $(\psi, \theta, \phi)$, and the coefficients $a_{lm}(\psi,\theta,\phi)$ in this new coordinate system can be calculated by
\begin{eqnarray}
a_{lm}(\psi,\theta,\phi)=\sum_{m'=-l}^{l} a_{lm'}D^{l}_{mm'}(\psi,\theta,\phi),
\label{almp}
\end{eqnarray}
where $a_{lm}\equiv a_{lm}(0,0,0)$ are the coefficients defined
in the Galactic coordinate system, and $D^{l}_{mm'}(\psi,\theta,\phi)$
is the Wigner rotation matrix \citep{Angular_Momentum}. Similar to Eq. (\ref{dpow}), we can
define the power spectrum $D(l;\psi,\theta,\phi)$. It is easy to find that $D(l;\psi,\theta,\phi)$ is
independent of the angle $\psi$, so in this paper we only
consider two Euler angle $\hat{\bf q}\equiv(\theta,\phi)$ and set
$\psi=0$. If we consider $\hat{\bf q}$ as a vector, which labels
the $z$-axis direction in the rotated coordinate system, then
$(\theta,\phi)$ is the polar coordinate of this direction in the Galactic system 
\footnote{Throughout this paper, we use the polar coordinate $(\theta,\phi)$ in the Galactic system, which relates to the Galactic coordinate ($l$, $b$) by $l=90^{\rm o}-\theta$ and $b=\phi$.}.

Now, we can define the rotationally variable parity
parameter $G(l;\hat{\bf q})$  by replacing $C(l)$ in Eq. (\ref{est}) with
$D(l;\hat{\bf q})$, and estimate the maxima
and minima of $G(l;\hat{\bf q})$ for different 
Euler angles $\hat{\bf q}$. By the definition, the parity
parameter $G(l;\hat{\bf q})$ depends on the coefficients
$a_{l0}(\hat{\bf q})$ as follows:
\begin{eqnarray}
 G(l;\hat{\mathbf q})
=\frac{P^+(l)-X^{+}(l;\hat{\bf q})}{P^-(l)-X^{-}(l;\hat{\bf q})},
\label{map}
\end{eqnarray}
where $X^{\pm}(l;\hat{\bf q})\equiv\frac{1}{4\pi}\sum_{l'=2}^{l} a^2_{l'0}(\hat{\mathbf q})\Gamma^{\pm}(l')$, and 
\begin{eqnarray}
a^2_{l0}(\hat{\bf q})&=&\sum_{mm'} a_{lm}a^*_{lm'} D^l_{0m}(\hat{\bf q}) {D^{l~*}_{0m'}}(\hat{\bf q}) \nonumber\\
&=&\frac{4\pi}{2l+1}\sum_{mm'} (-1)^{m+m'}a_{lm}a^*_{lm'} Y^*_{lm}(\hat{\bf q}) {Y_{lm'}}(\hat{\bf q}). \nonumber
\end{eqnarray}
So the cross-term $a_{lm}a^*_{lm'}$ is responsible for the angular dependency of the parity parameter $G(l;\hat{\mathbf q})$. We can also calculate the
difference between $G(l;\hat{\mathbf q})$ and $g(l)$ by
\begin{eqnarray}
\frac{G(l;\hat{\mathbf q})-g(l)}{g(l)}
\simeq \frac{X^{-}(l;\hat{\bf q})-X^+(l;\hat{\bf q})/g(l)}{P^-(l)}.
\label{map2}
\end{eqnarray}
From the relation $\Delta(\l)=O(\frac{1}{2l})$, we know that $X^{\pm}(l;\hat{\bf q})\ll P^{-}(l)$, and $\frac{G(l;\hat{\mathbf q})-g(l)}{g(l)}\ll1$ for $l>3$. 
So, we conclude that $G(l;\hat{\mathbf q})$ mainly stands for the amplitude of the original parity parameter $g(l)$. At the same time, 
due to the rotational variance of $G(l;\hat{\mathbf q})$, we can study the possible preferred direction, which may reveal hints on the origin of the observed parity asymmetry in CMB field.

\begin{figure}
  \begin{center}
\hbox{
    \centerline{\epsscale{.40}\plotone{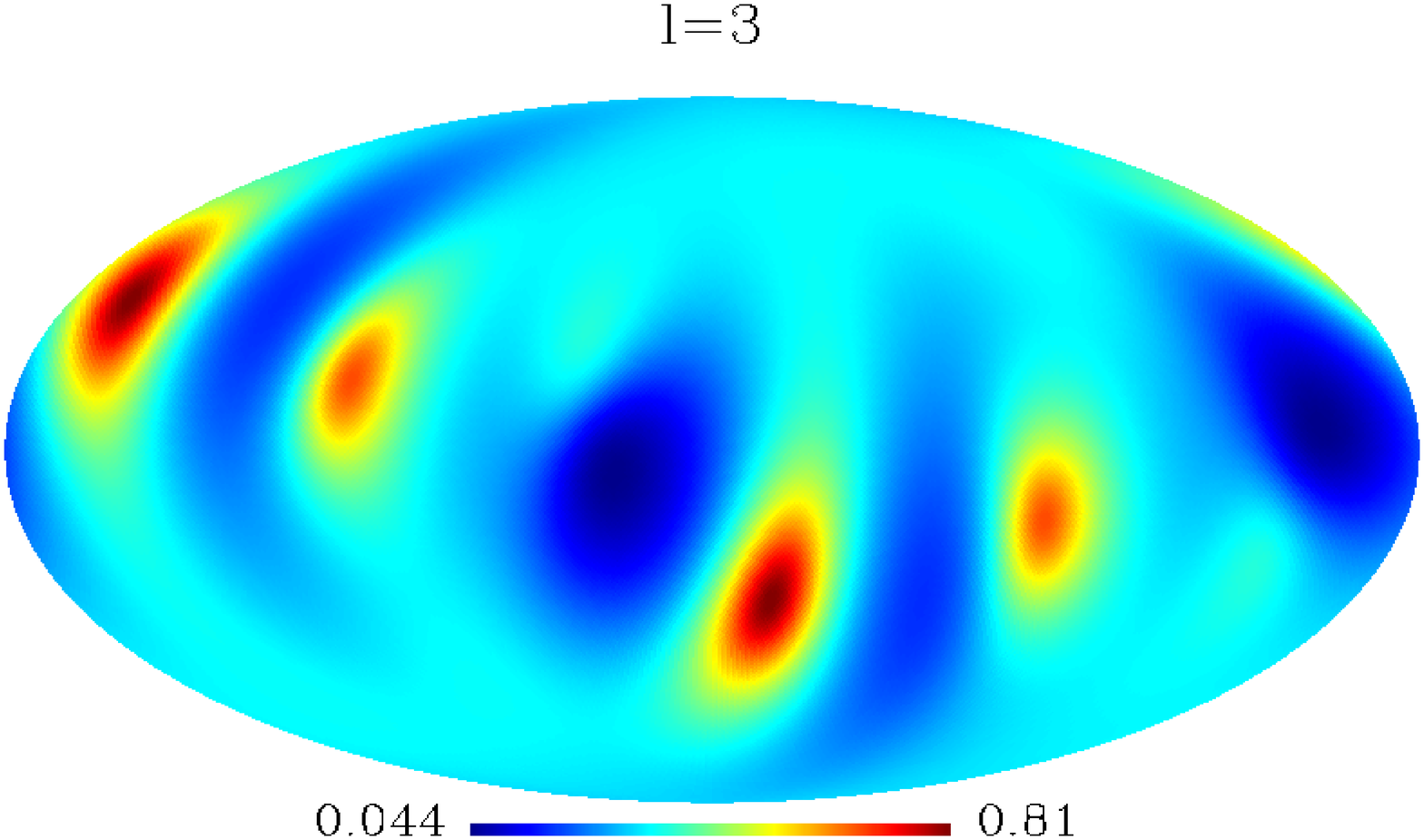}\epsscale{.40}\plotone{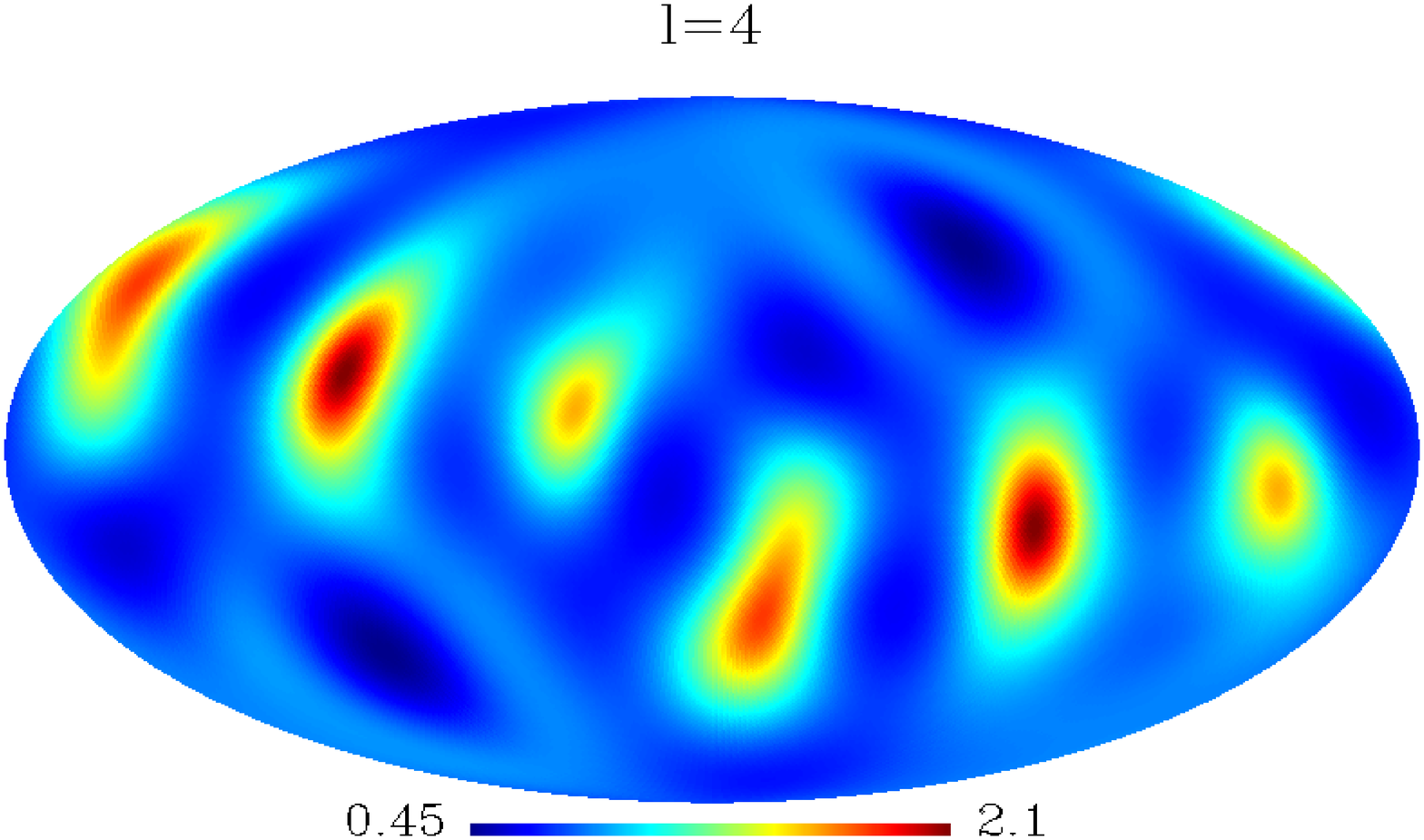}}}
\hbox{
    \centerline{\epsscale{.40}\plotone{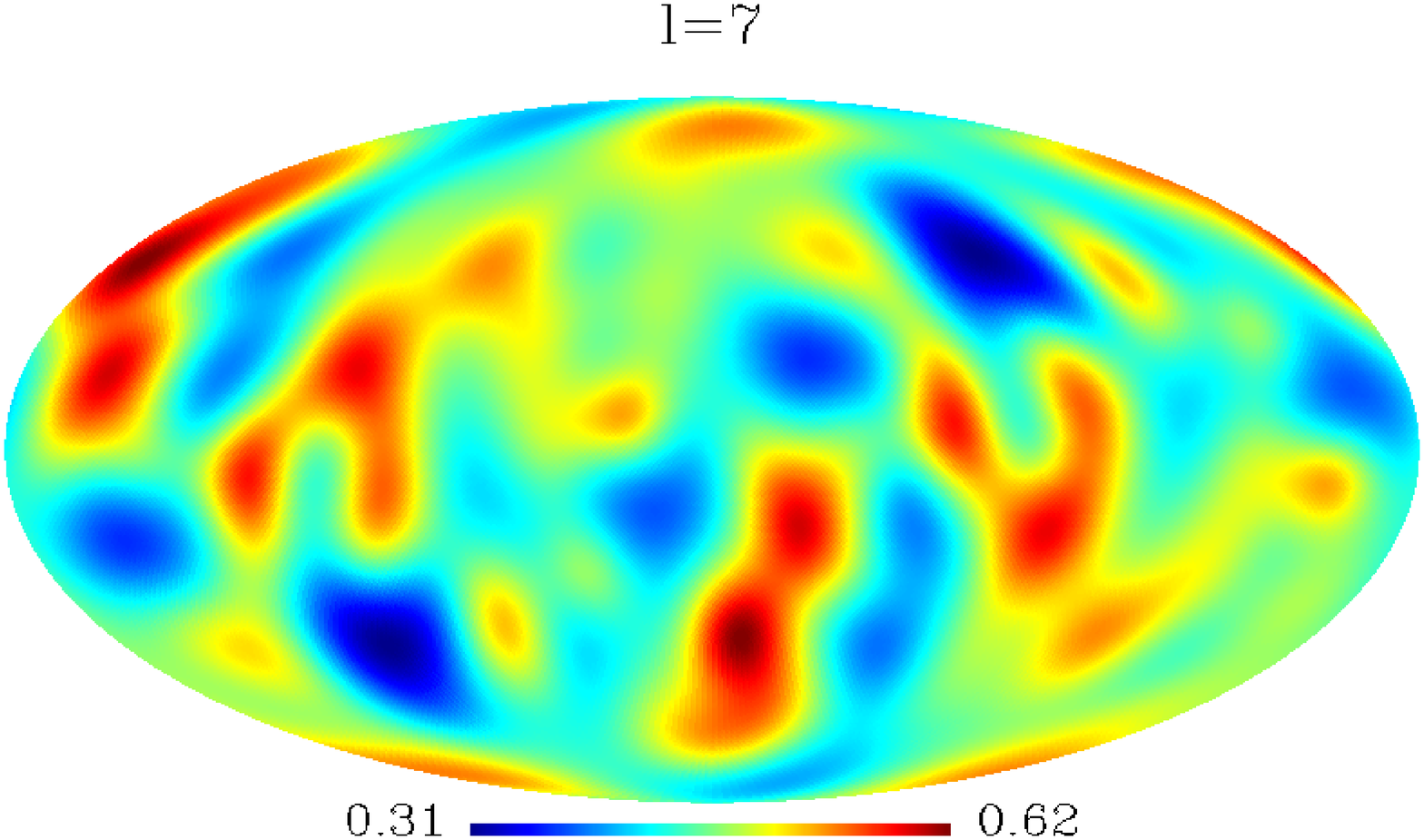}\epsscale{.40}\plotone{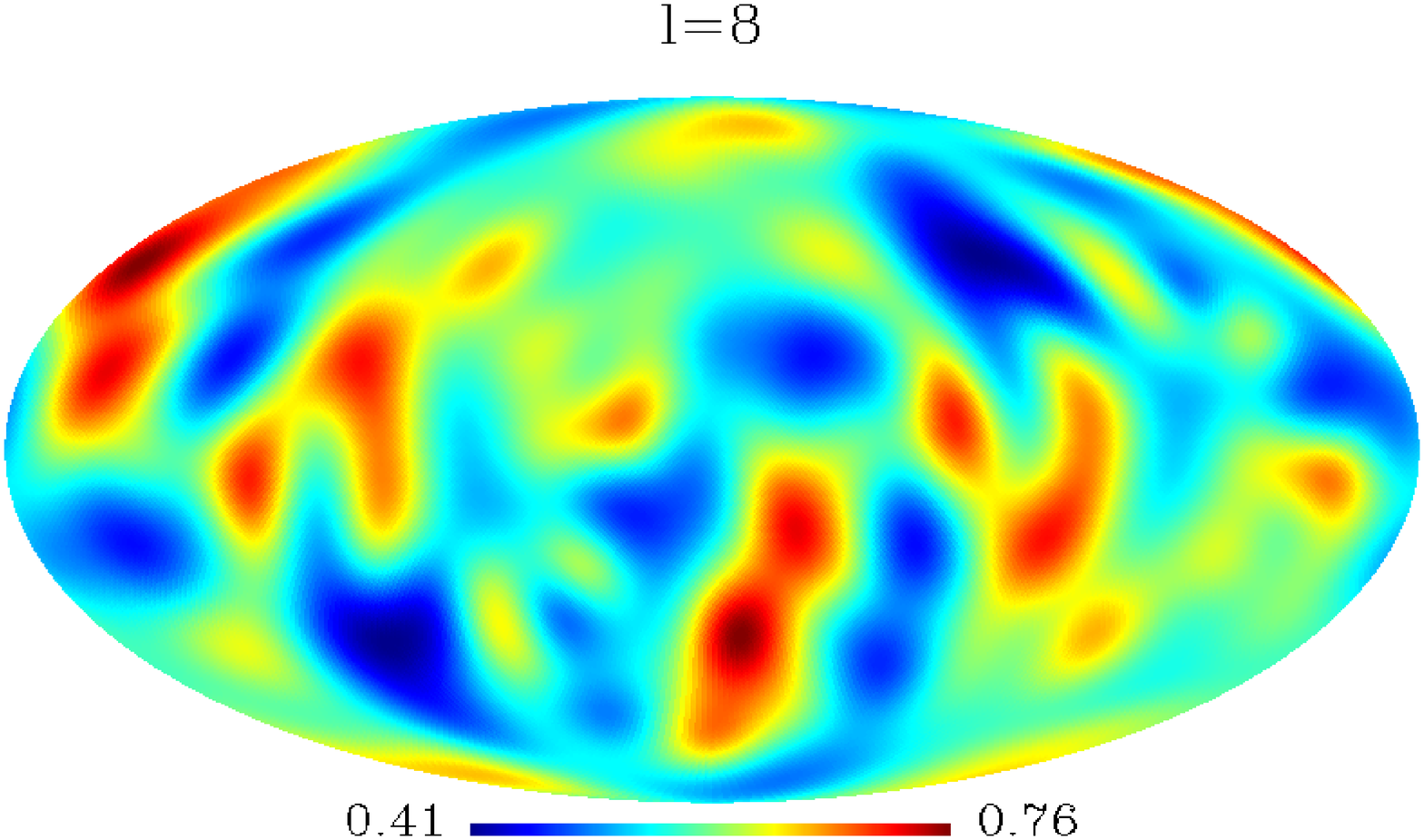}}}
\hbox{
    \centerline{\epsscale{.40}\plotone{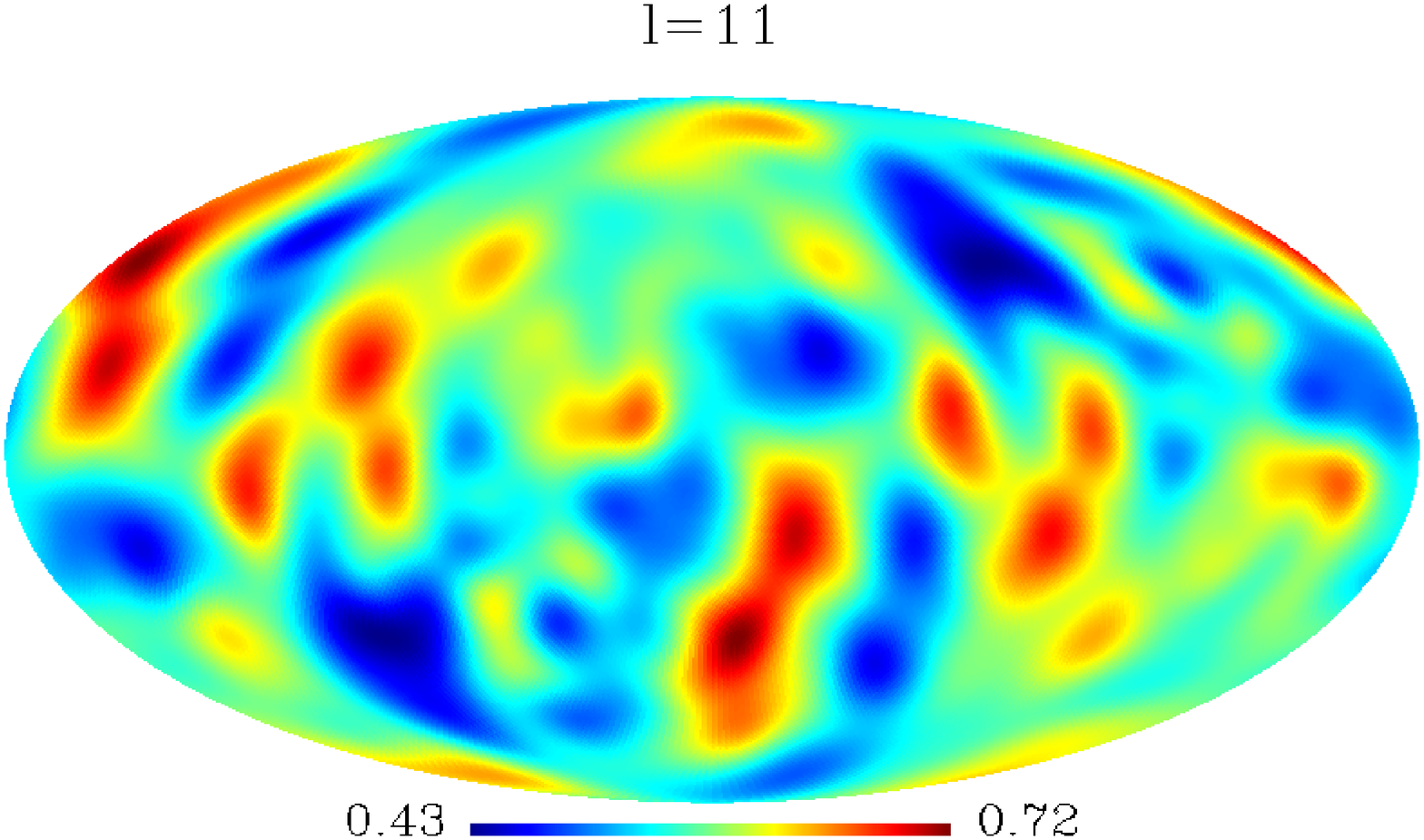}\epsscale{.40}\plotone{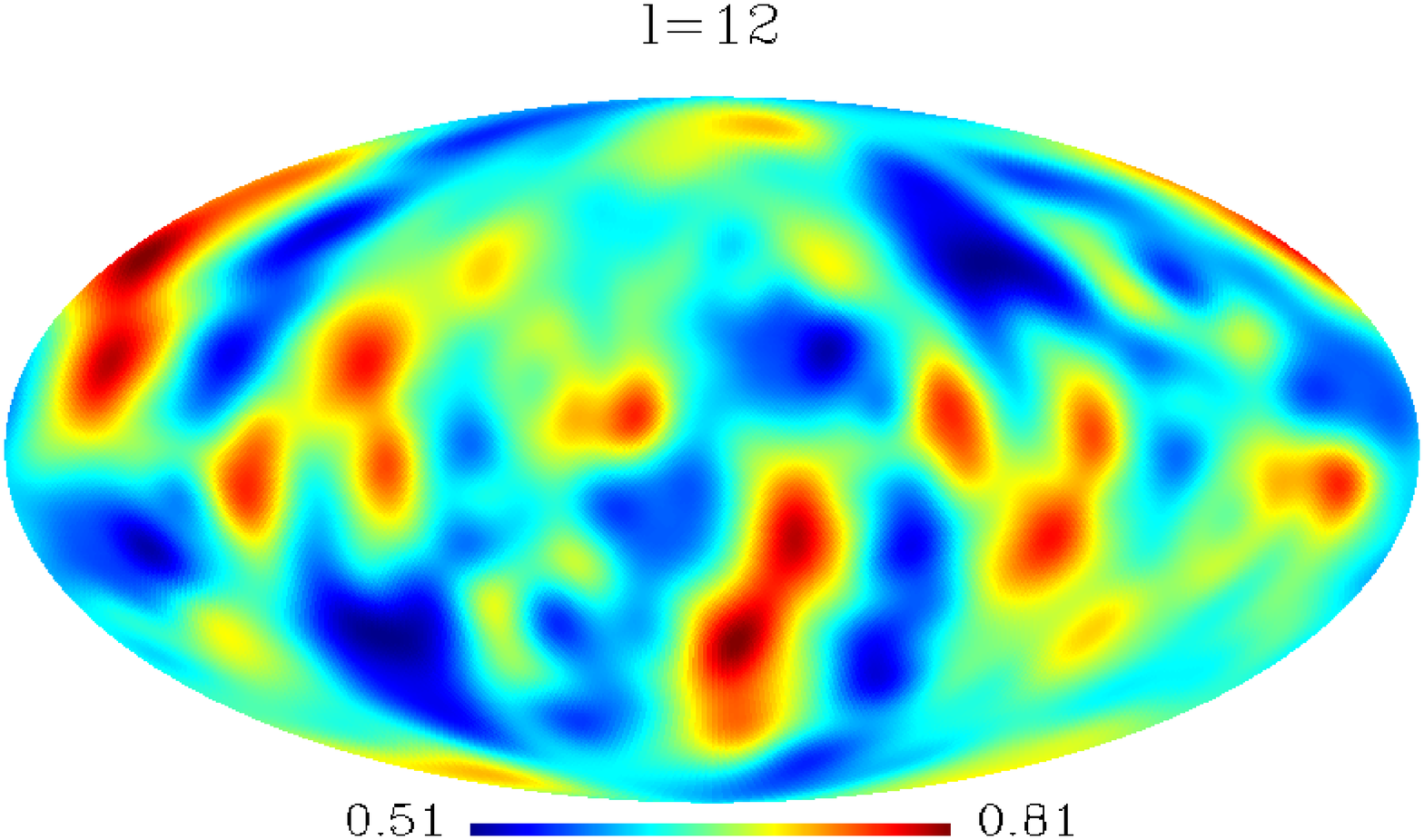}}}
\hbox{
    \centerline{\epsscale{.40}\plotone{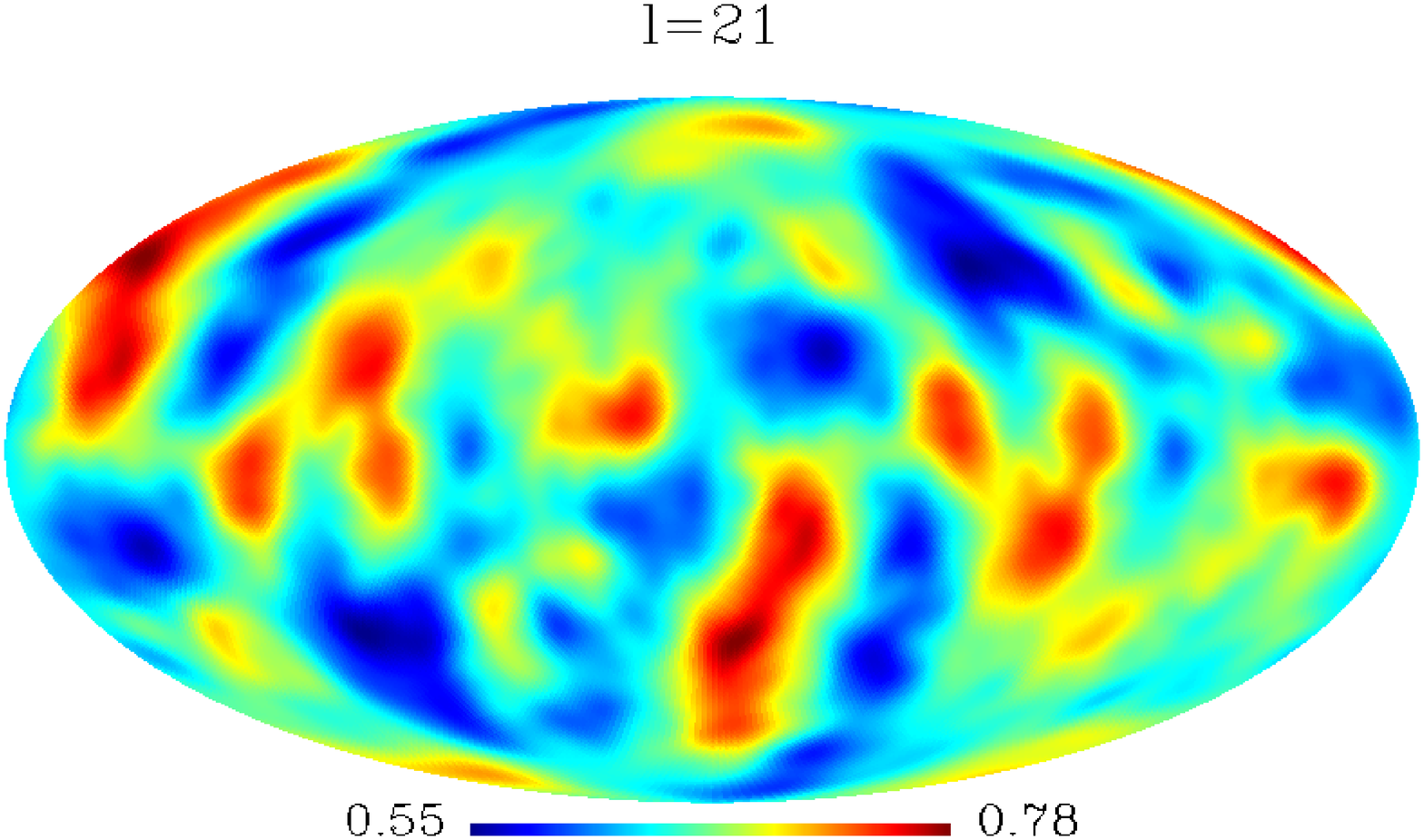}\epsscale{.40}\plotone{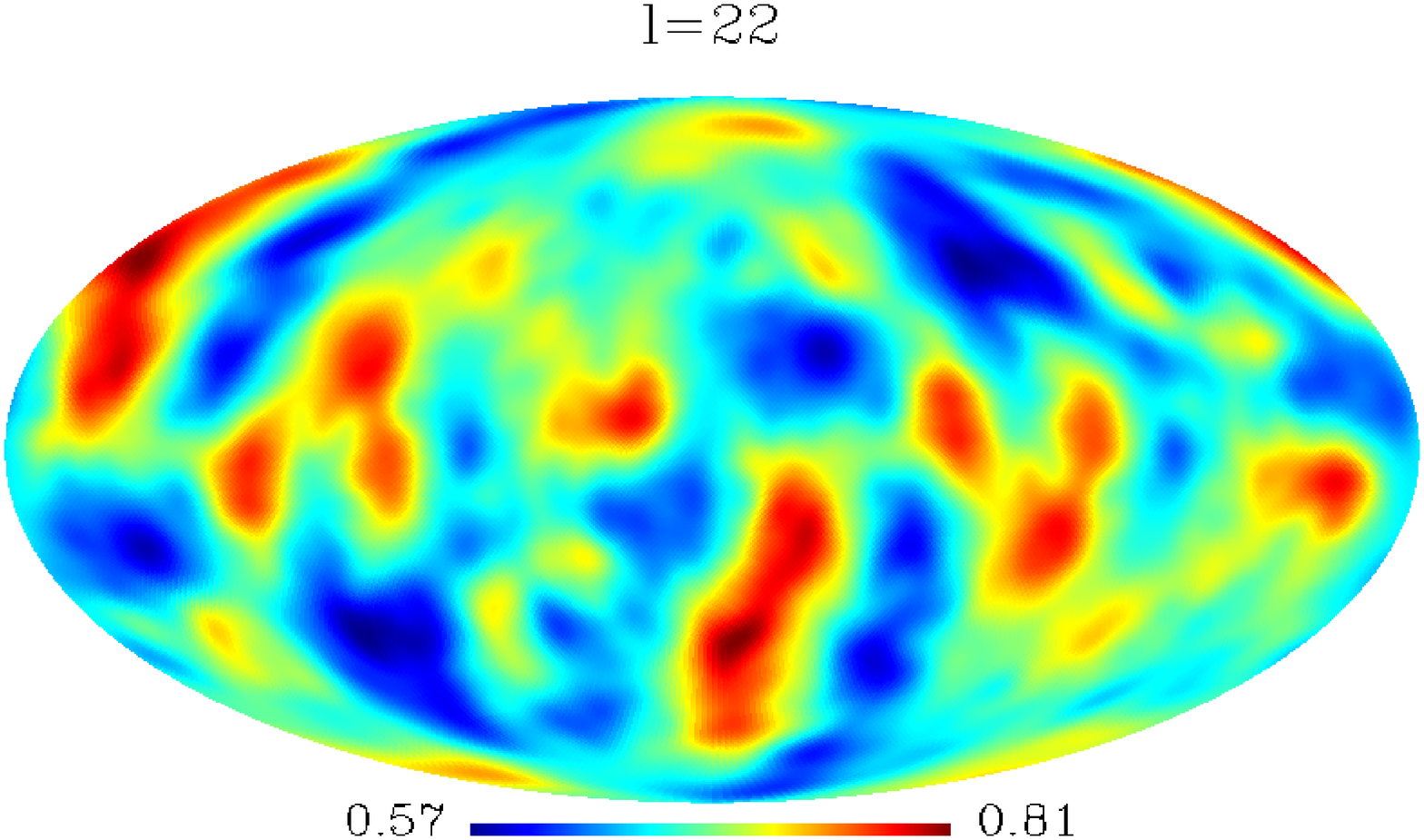}}}
    \caption{The parameter $G(l;\hat{\bf q})$ based on the estimators in Eq. (\ref{dpow}) as a function of $\hat{\bf q}\equiv (\theta,
    \phi)$ for $l=3,~7,~11,~21$ (left panels) and $l=4,~8,~12,~22$ (right panels).}
    \label{Glq}
  \end{center}
\end{figure}

\begin{figure}
\epsscale{.80}
\plotone{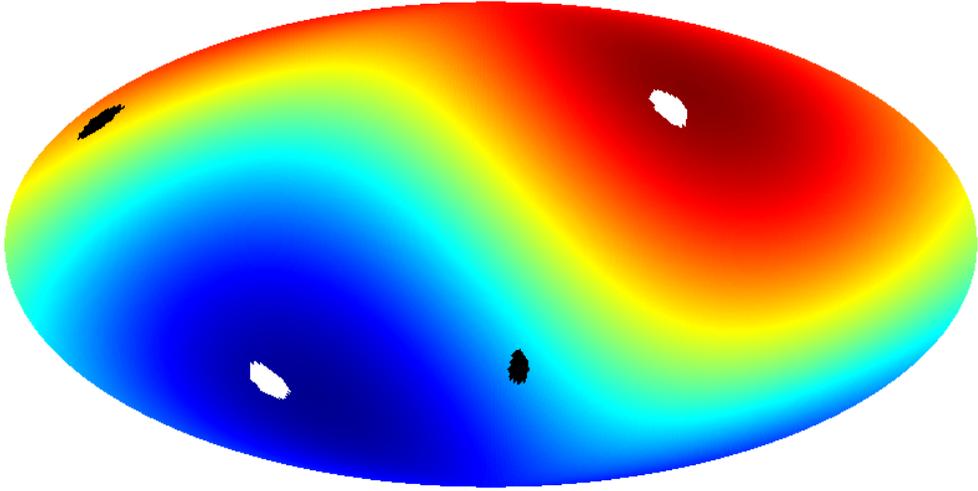}
\caption{The ILC7 dipole component in the Galactic coordinate system.
The functions $G(l;\hat{\bf q})$ ($4 \le l \le 22$) minimize at the
white regions, and maximize at the black regions. Note that the
center direction of the white regions are $\hat{\bf q}=(46.59^{\rm
o},277.98^{\rm o})$ and $-\hat{\bf q}$, while those of the black
regions are $\hat{\bf q}=(50.55^{\rm o},167.34^{\rm o})$ and
$-\hat{\bf q}$. }.\label{dip7}
\end{figure}

Let us show that $G(l;\hat{\mathbf q})$ map depends on the angular
$\hat{\mathbf q}$. As we have mentioned, $\hat{\bf q}$ labels the
$z$-axis direction in the rotated coordinate system, and
$(\theta,\phi)$ is just the polar coordinate of this direction in Galactic coordinate.
We plotted the parameter $G(l;\hat{\mathbf q})$ as a function of
$\hat{\mathbf q}$ for $3\le l \le 22$, and found that
$G(l;\hat{\mathbf q})$ have the similar morphology for
$l \ge 4$, which is clearly shown in Fig. \ref{Glq}. (Note that, the morphology of $l=3$ map is different, which may relate to the unsolved low quadrupole problem 
as well as the alignment of quadrupole and octupole \citep{wmap_anomly}.)
In Table \ref{table1}, we list the preferred directions $\hat{\bf q}$, where the parity parameter $G(l;\hat{\bf q})$ for each $l$ is minimized (note that different from the problem in \citep{aaaaa}, 
here as the widely discussed aligment problem of quadrupole and octupole in \citep{dt2},
the uncertainties of the preferred directions are difficult to be defined), which are very close with each other for $l\ge 4$.
So, we can choose the special direction $\hat{\bf q}$ (note that $-\hat{\bf q}$ is another equivalent preferred direction), where all the parameters $G(l;\hat{\mathbf q})$ are minimized or maximized. 
We picked out these regions, and plotted them in the Galactic coordinate system in
Fig. \ref{dip7}. It is interesting to find that the preferred directions
$\hat{\bf q}$, where parity violation is largest, are coincident
with the WMAP7 KD direction \citep{wmap7_dipole}, while the preferred directions
$\hat{\bf q}$, where parity asymmetry is smallest, are nearly perpendicular to the KD direction. 
If we assume the parity asymmetry in the CMB has the cosmological origin, it is very hard to explain these 
coincidences. So, the coincidence of the preferred direction $\hat{\bf q}$ with the WMAP7 KD direction implies that the CMB parity 
asymmetry may relate to the possible contamination of residual WMAP KD component.

Although we will not detailedly study the physical mechanism in this paper, we could provide some possible explanations for this coincidence problem.
It is noticed that there is not a great deal of residual dipole in the WMAP data, so a possible explanation could be
connected to the use of the dipole as a photemetric calibrator in the WMAP data set.

Another possible explanation is related to the contaminations generated by 
the collective emission of Kuiper Belt Objects (KBOs) and other minor bodies in the solar system where the KD direction is localized.
Since the emission of KBOs is nearly independent of the frequency in the WMAP frequency range, this contamination is very hard to be removed in the WMAP data analysis. 
In \citep{maris,kbo2}, it was discussed that this foreground residual could leave significant parity asymmetry in the CMB data.  

Besides, the explanation may also relate to the measure deviation of the WMAP kinematic dipole, which could be caused by 
the measure error in dipole direction, antenna pointing direction, sidelobe pickup contamination, and so on. 
In \citep{liu2011}, it was found that this KD deviation could generate the artificial CMB anisotropies in the low multipoles. 
If this is true, these artificial components may account for the CMB parity violation.

\begin{figure}
  \begin{center}
\hbox{
    \centerline{\epsscale{.40}\plotone{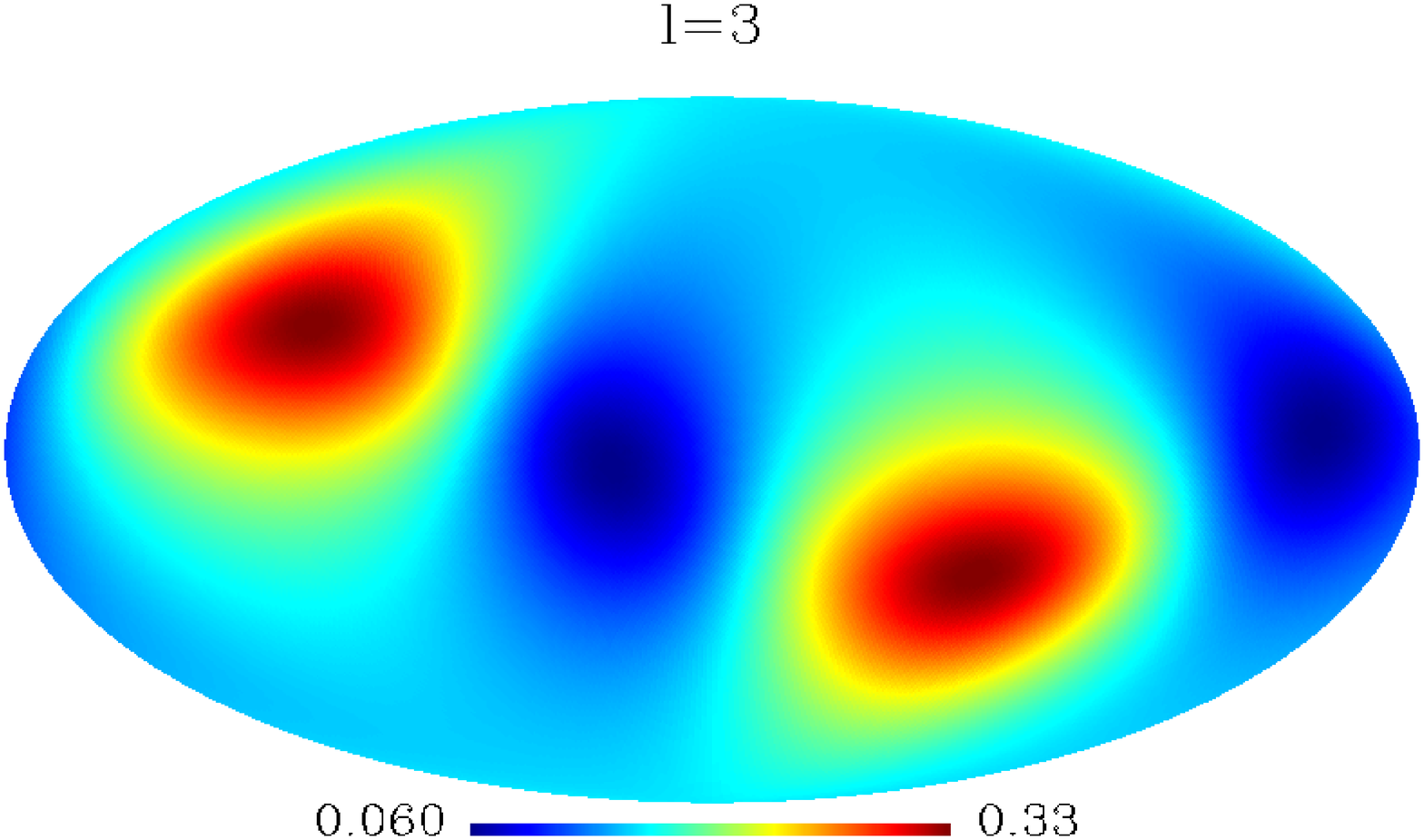}\epsscale{.40}\plotone{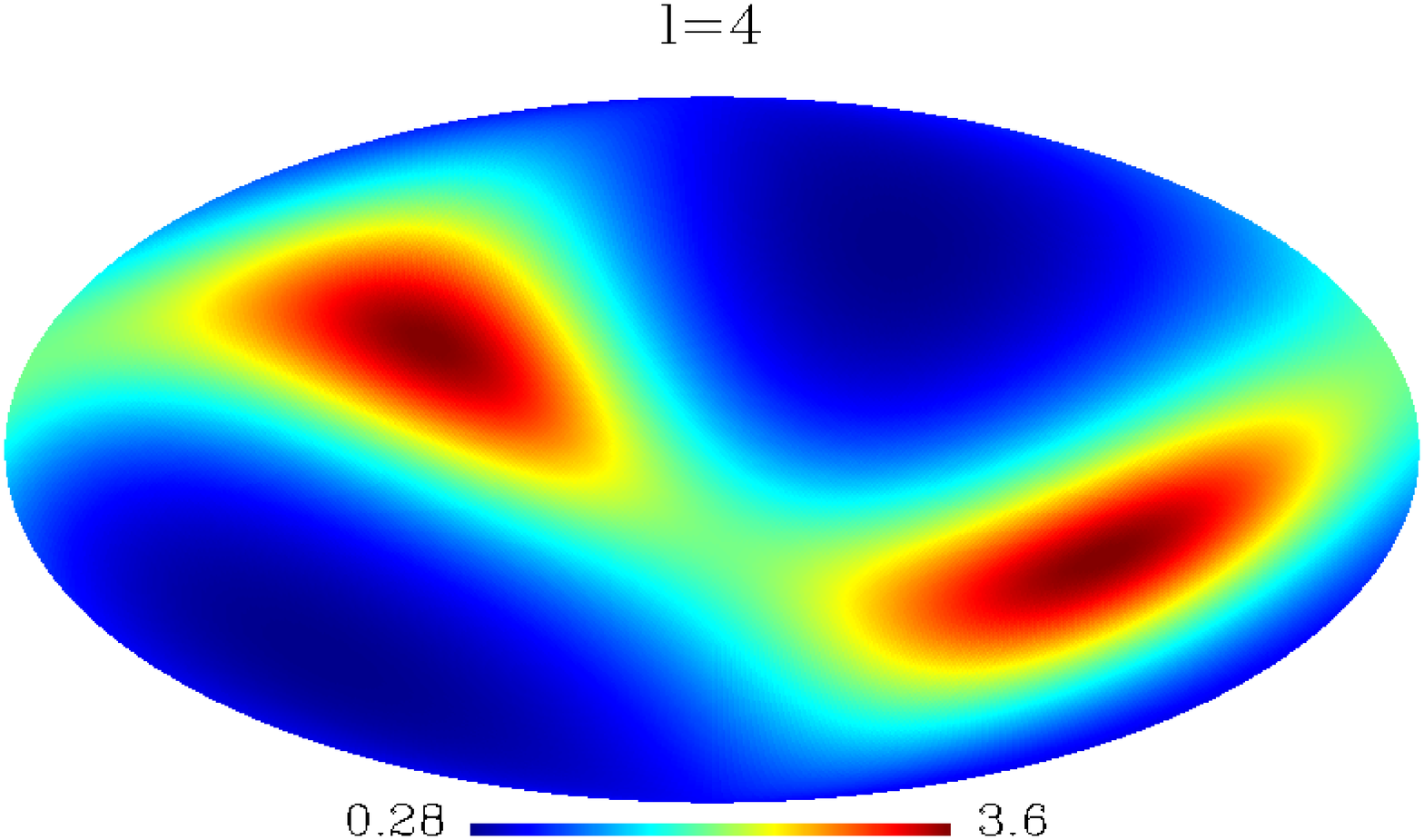}}}
\hbox{
    \centerline{\epsscale{.40}\plotone{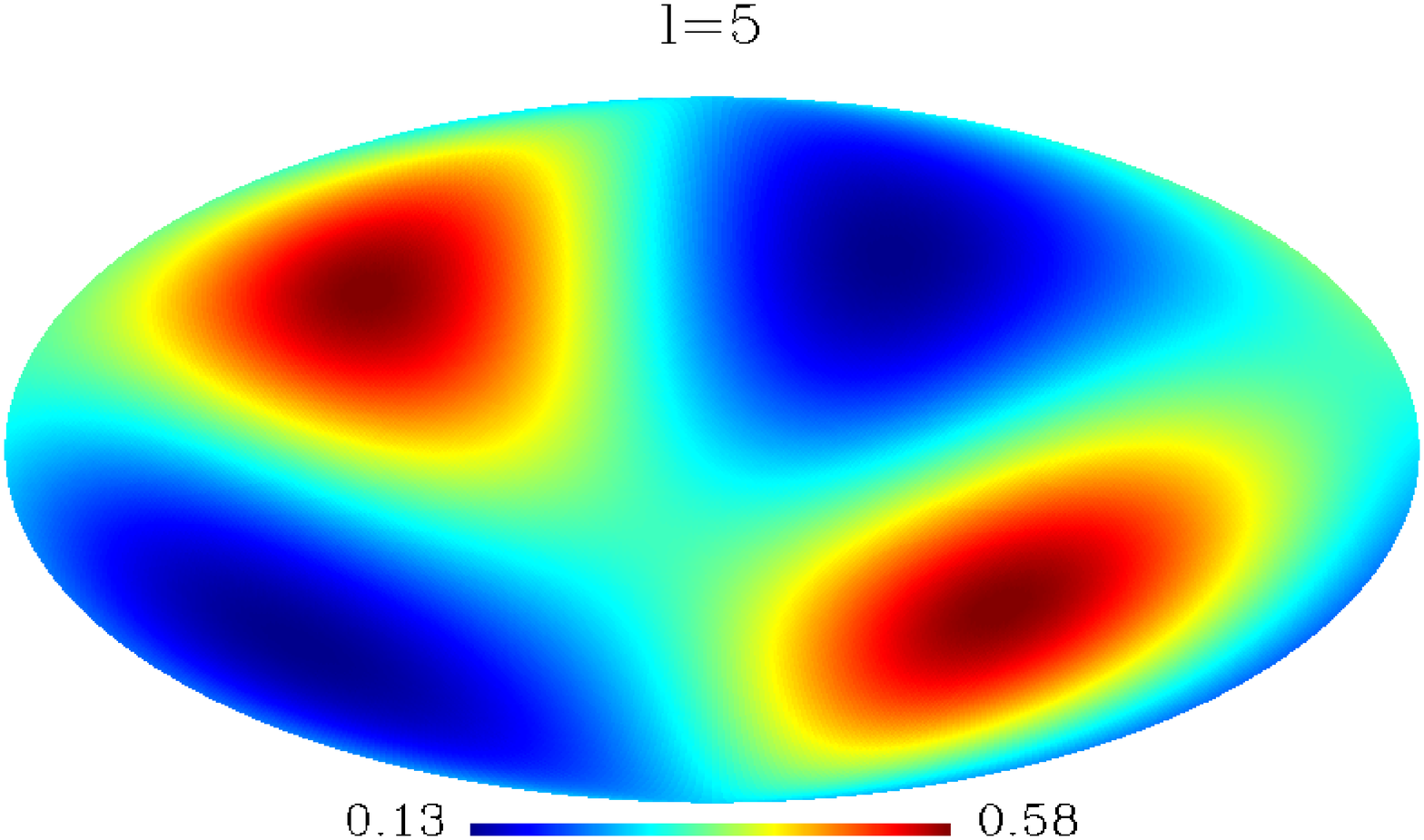}\epsscale{.40}\plotone{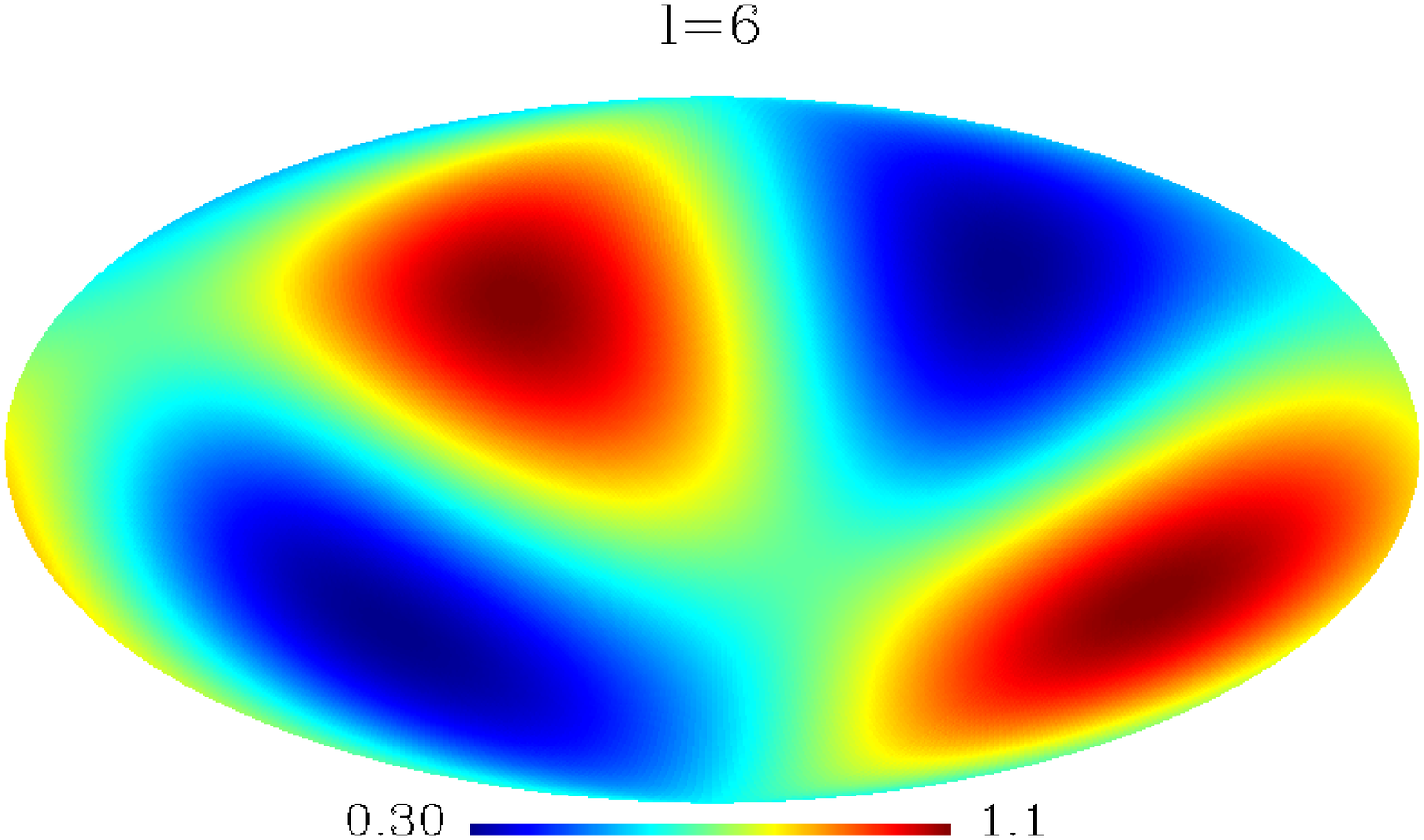}}}
\hbox{
    \centerline{\epsscale{.40}\plotone{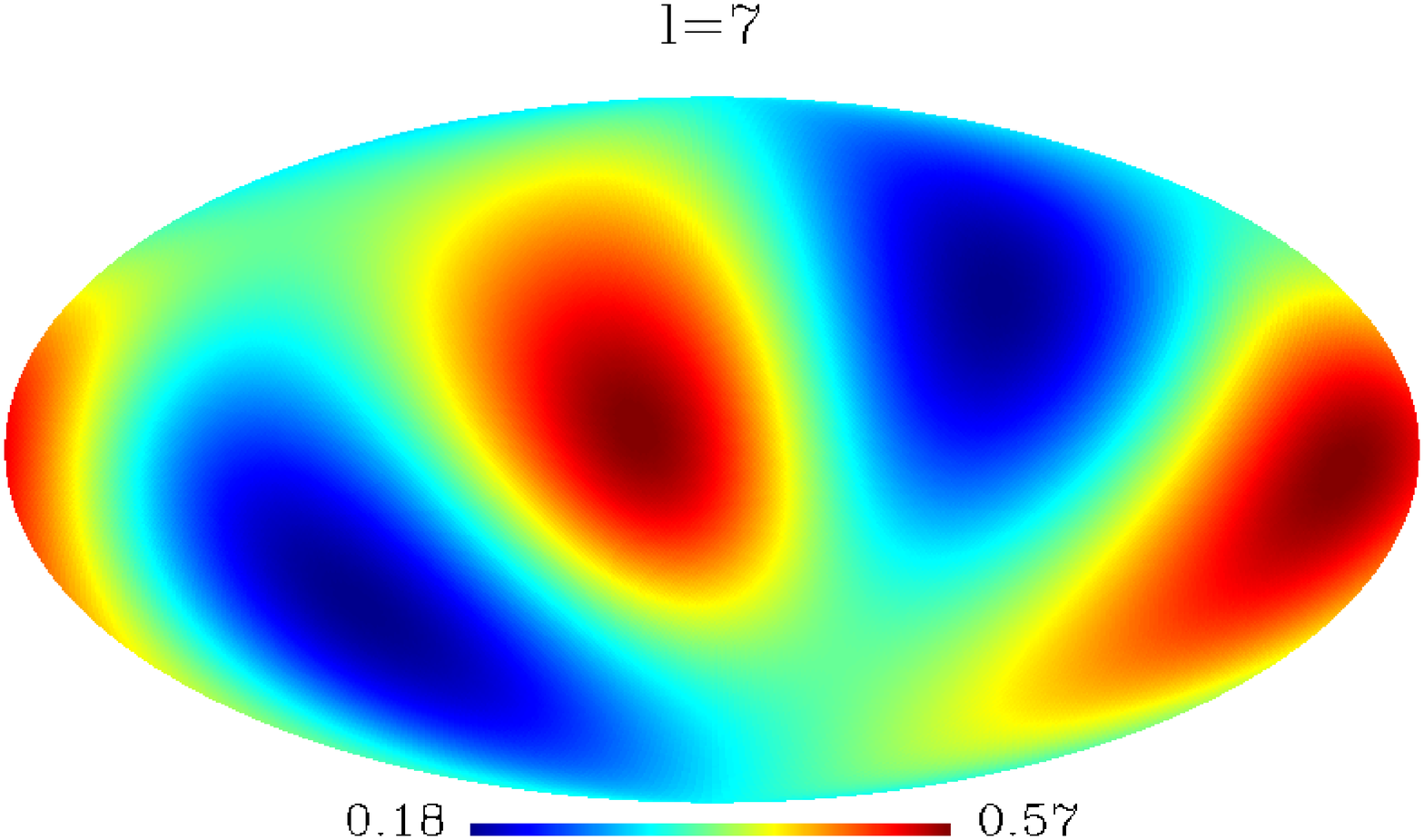}\epsscale{.40}\plotone{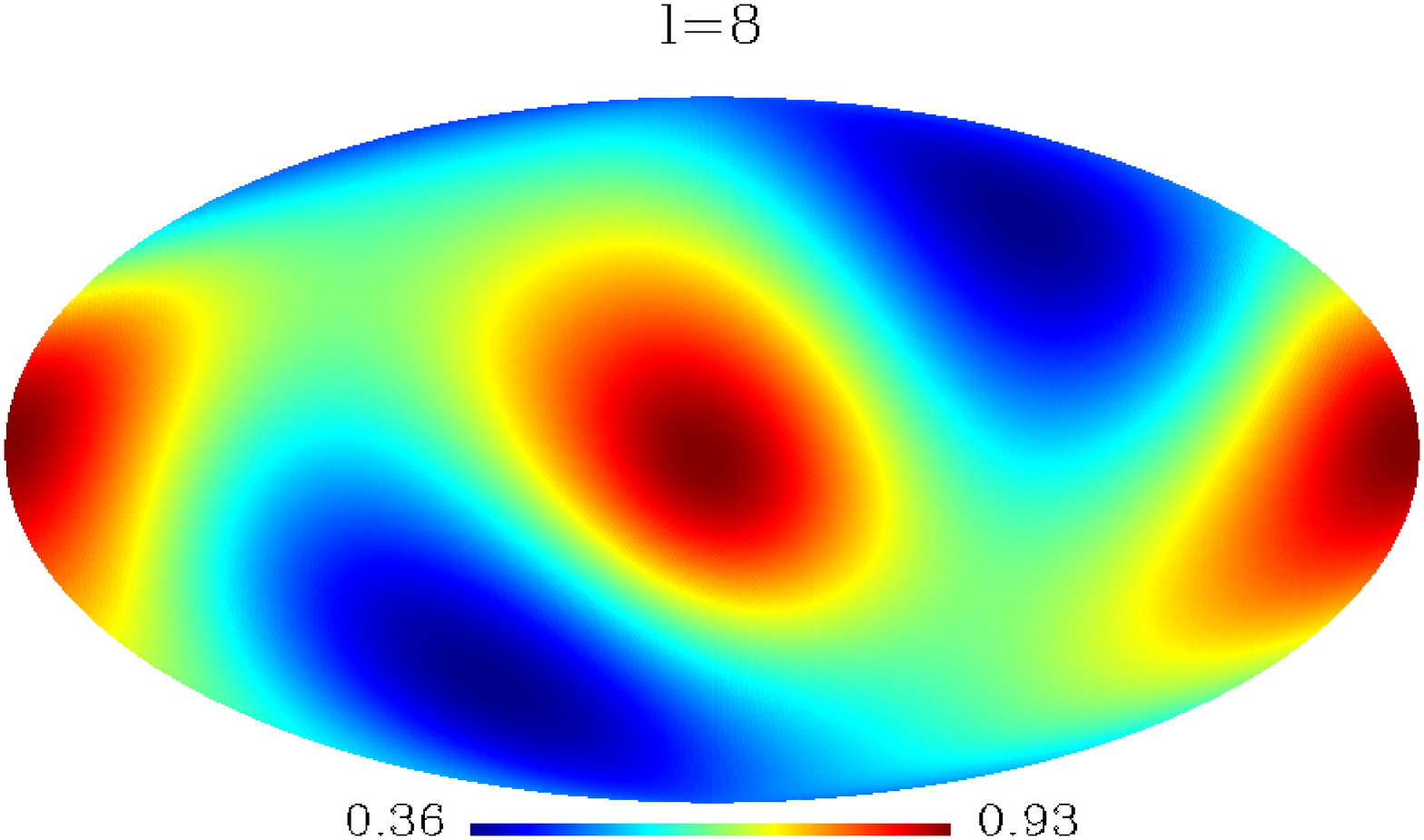}}}
\hbox{
    \centerline{\epsscale{.40}\plotone{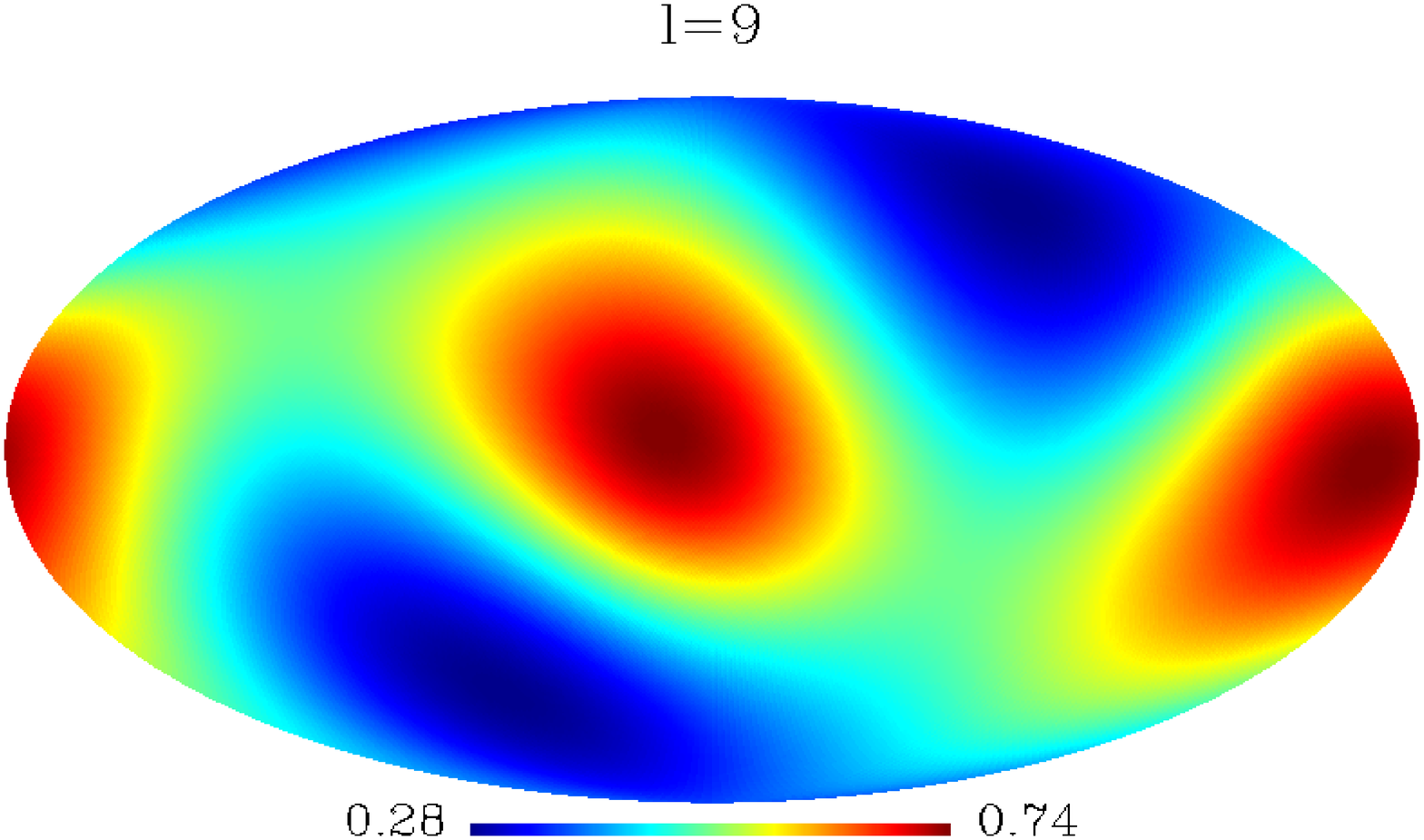}\epsscale{.40}\plotone{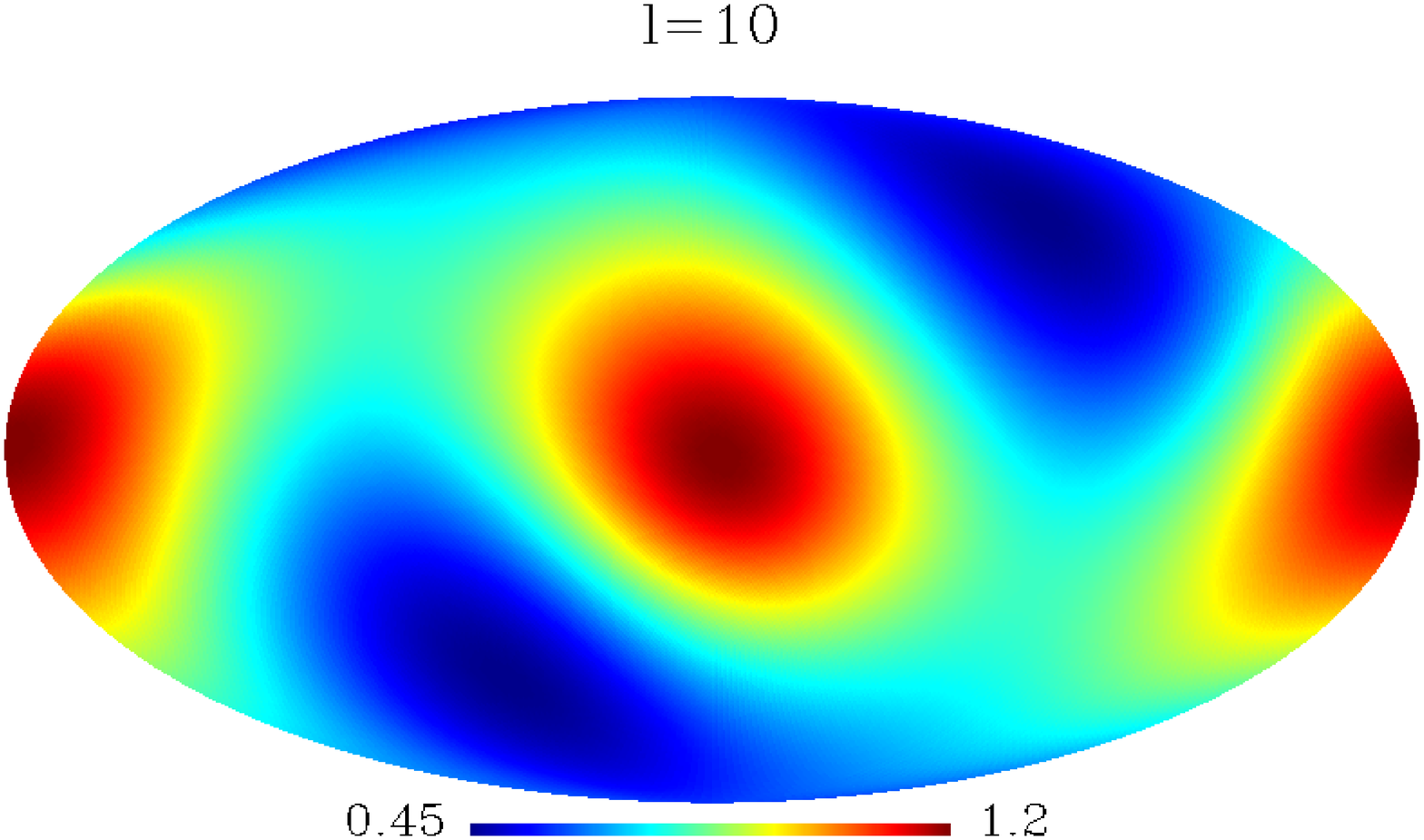}}}
    \caption{The parameter $\tilde{G}(l;\hat{\bf q})$ based on the estimators in Eq.(\ref{tegmark}) as a function of $\hat{\bf q}\equiv (\theta,
    \phi)$ for $l=3,~5,~7,~9$ (left panels) and $l=4,~6,~8,~10$ (right panels).}
    \label{GGlq}
  \end{center}
\end{figure}

In order to cross-check this result, we consider another rotationally variant estimator, which is proposed by \citep{dt2}
\begin{eqnarray}
\tilde{D}(l)\equiv\frac{1}{2l+1}\sum_{m=-l}^{l} m^2 |a_{lm}|^2.
\label{tegmark}
\end{eqnarray}
If our Universe is statistically isotropic, the ensemble average of this estimator is related to the power spectrum as follows:
\begin{eqnarray}
\langle \tilde{D}(l)\rangle= \frac{l(l+1)}{3} \langle C_l \rangle.
\end{eqnarray}
As well discussed, this statistic has also chosen the preferred direction, i.e. the $z$-axis direction. In addition, this statistic favors high $m$s and so it works well in searches for planarity. In a quantum mechanical system, this quantity also corresponds to the angular momentum along the $z$-axis direction \citep{Angular_Momentum}.
Due to the rotational variance of this quantity, we can define $\tilde{D}(l;\hat{\bf q})$ and the corresponding parity parameter $\tilde{G}(l;\hat{\bf q})$, where $\hat{\bf q}$ is the Euler rotation angle. We notice that in the definition of $\tilde{G}(l;\hat{\bf q})$, due to the factor $m^2$ in $\tilde{D}(l)$, the weight of higher multipoles are much greater than the lower ones. So in this paper, we only consider the parity parameter $\tilde{G}(l;\hat{\bf q})$ for $l\le 10$, where the parity violation is obvious. We plot the quantity $\tilde{G}(l;\hat{\bf q})$ as a function of $\hat{\bf q}$ for low multipoles in Fig. \ref{GGlq}, and find that these maps have the quite similar morphology, especially for $4\le l \le 10$. The preferred directions $\hat{\bf q}$ (similar, $-\hat{\bf q}$ is another equivalent preferred direction), where the parity parameter $\tilde{G}(l;\hat{\bf q})$ for each $l$ is minimized, are also listed in Table \ref{table2}.
Again, we find the similar results: for $4 \le l \le 10$, the bluer regions (larger parity violation) are excellently coincident to the WMAP7 KD direction, while the redder regions (smaller parity asymmetry) are nearly perpendicular to the KD direction. So this cross-check agrees with our previous finding: the CMB parity asymmetry in the low multipoles may connect with the possible contamination of WMAP KD component.

\section{Conclusion}

In this paper, we have investigated the directional properties of CMB parity asymmetry. In order to break the rotational invariance of the CMB power spectrum, 
we defined two different power spectrum estimators so that a special direction is picked out. By rotating these estimators with respect to the Galactic coordinate system, 
we studied the corresponding parity parameters as functions of the preferred directions $\hat{\bf q}$, where the parity parameters are minimized or maximized. 
We found that these preferred directions are aligned (parallel or perpendicular) with the WMAP7 kinematic dipole direction, which implies that the CMB parity asymmetry may be produced by the systematics associated with kinematic dipole. This study also shows that the effect of the WMAP kinematic dipole may extend to the higher multipoles $l\sim 22$. 

The Planck surveyor possesses wide frequency coverage and systematics distinct from the WMAP.
In particular, it may take advantage of both COBE and WMAP results for the dipole calibration and meanwhile Planck satellite will have high signal-to-noise ratio in polarization data.
Therefore, we may apply the similar tests on the CMB TE and EE data from the Planck surveyor, and hope to resolve the association between the parity asymmetry and kinematic dipole.



\acknowledgments
We appreciate useful discussions with P. Coles.
We acknowledge the use of the Legacy Archive for Microwave Background Data Analysis (LAMBDA).
Our data analysis made the use of HEALPix \citep{healpix} and GLESP \citep{glesp}.
This work is supported in part by Danmarks Grundforskningsfond, which allowed the establishment of the Danish Discovery Center.
This work is supported by FNU grant 272-06-0417, 272-07-0528 and 21-04-0355.

\clearpage



\clearpage

\clearpage


\clearpage

\clearpage

\clearpage

\begin{table}
\begin{center}
\caption{The WMAP7 kinematic dipole direction is compared with the preferred direction $\hat{\bf q}=(\theta,\phi)$, where the parity parameter $G(l;\hat{\bf q})$ (based on the estimator in Eq. (\ref{dpow})) is minimized. Note that $-\hat{\bf q}$ is another preferred direction. \label{table1}}
\begin{tabular}{crrrr}
\tableline\tableline
     & {$\theta$ [$^{\rm o}$]} & {$\phi$ [$^{\rm o}$]} & {$\cos\alpha$\tablenotemark{a}}\\
\tableline \\
KD     &  41.74    &  263.99   &  ------    \\
$l=3$  &  85.22    &  204.61   &  0.400     \\
$l=4$  &  46.59    &  280.89   &  0.975     \\
$l=7$  &  48.19    &  279.14   &  0.976     \\
$l=8$  &  48.99    &  277.03   &  0.979     \\
$l=11$ &  49.77    &  277.73   &  0.976     \\
$l=12$ &  49.77    &  277.73   &  0.976     \\
$l=21$ &  51.32    &  283.36   &  0.957     \\
$l=22$ &  50.50    &  284.06   &  0.957     \\
\tableline
\end{tabular}
\tablenotetext{a}{$\alpha$ is the angle between $\hat{\bf q}$ and the KD direction.}
\end{center}
\end{table}

\clearpage

\begin{table}
\begin{center}
\caption{The WMAP7 kinematic dipole direction is compared with the preferred direction $\hat{\bf q}=(\theta,\phi)$, where the parity parameter $\tilde{G}(l;\hat{\bf q})$ (based on the estimator in Eq. (\ref{tegmark})) is minimized. Note that $-\hat{\bf q}$ is another preferred direction. \label{table2}}
\begin{tabular}{crrrr}
\tableline\tableline
     & {$\theta$ [$^{\rm o}$]} & {$\phi$ [$^{\rm o}$]} & {$\cos\alpha$\tablenotemark{a}}\\
\tableline \\
KD     &  41.74    &  263.99   &  ------    \\
$l=3$  &  86.42    &  206.02   &  0.401     \\
$l=4$  &  45.80    &  303.20   &  0.890     \\
$l=5$  &  48.19    &  305.86   &  0.867     \\
$l=6$  &  52.08    &  274.22   &  0.975     \\
$l=7$  &  57.91    &  279.84   &  0.939     \\
$l=8$  &  39.20    &  255.57   &  0.994     \\
$l=9$  &  37.20    &  252.90   &  0.989     \\
$l=10$ &  40.30    &  249.17   &  0.985     \\
\tableline
\end{tabular}
\tablenotetext{a}{$\alpha$ is the angle between $\hat{\bf q}$ and the KD direction.}
\end{center}
\end{table}

\clearpage



\clearpage




\end{document}